\documentclass[11pt, a4paper]{article}

\usepackage{latexsym}
\usepackage{graphicx}
\usepackage{cite}
\usepackage{amsmath}
\usepackage{amssymb}
\begin{document}
\title{My Understanding for Static and Dynamic Light Scattering}
\author{Yong Sun{\footnote{Email: ysun200611@yahoo.ca}}\\
\emph{ysun200611@yahoo.ca}
\\\emph{Burnaby, BC, Canada}}
\maketitle

\begin{abstract}
The results obtained using my computing program are consistent with the values obtained twenty years ago.
It also makes me believe that how to obtain particle size information using the Light Scattering 
technique needs to be reconsidered.

\noindent  Static Light Scattering (SLS) and Dynamic Light Scattering (DLS)
are very important techniques to study the characteristics of 
nano-particles in dispersion. The data of SLS is determined by the optical characteristic
and the measured values of DLS are determined by optical 
and hydrodynamic characteristics of different size nano-particles in dispersion.
Then considering the optical characteristic of nano-particles and using the SLS technique further, 
the size distribution can be measured accurately and is also consistent with the results measured using 
the TEM technique. Based on the size distribution obtained using the SLS or TEM technique and the 
relation between the static and hydrodynamic radii, all the expected and measured values of 
$g^{\left( 2\right) }\left( \tau \right) $ investigated are very well consistent. 
Since the data measured using the DLS technique contains the information of the optical and hydrodynamic
properties of nano-particles together, 
therefore the accurate size distribution cannot be obtained from the experimental data of
 $g^{\left( 2\right) }\left( \tau \right) $ for an unknown sample.
The traditional particle information: apparent hydrodynamic radius and polydispersity index measured using the DLS technique are determined by the optical and hydrodynamic characteristics and size distribution together.
They cannot represent a number distribution of nano-particles in dispersion. 
Using the light scattering technique not only can measure the size distribution accurately but also can provide a method to understand the optical and hydrodynamic characteristics of nano-particles.

\end{abstract}

\section{INTRODUCTION}

Light Scattering has been considered to be a well established
technique and applied in physics, chemistry, biology, etc. as an
essential tool to investigate the characteristics of nano-particles
in dispersion. It consists of two parts: Static Light Scattering (SLS)
and Dynamic Light Scattering (DLS) techniques. The SLS technique measures the relation between
the scattering angles and scattered light intensity and then
simplify the relation to the Zimm plot, Berry plot or Guinier
plot etc. to get the root mean-square radius of gyration
$\left\langle R_{g}^{2}\right\rangle ^{1/2}$ and the molar mass of
nano-particles provided that the particle sizes are small\cite{re1,re2}.
The DLS technique has been considered that it can provide a fast and accurate
method to measure the particle sizes in dispersion. The apparent 
hydrodynamic radius and polydispersity index are obtained
using the method of Laplace transformation or
Cumulant analysis at a given scattering angle\cite{re3, re4, re5,
re6, re7}. Some researchers believe that the results measured using the SLS and
DLS techniques together can be used to obtain much more information of nano-particles
\cite{re8, re9, re10, re11, re12}. Although the theoretical function of the
normalized time electric field auto-correlation function of the scattered light $
g^{\left(1\right)}(\tau )$\cite{re4, re6, re13} has been proposed, the comparison
of the expected values and experimental data of the
normalized time auto-correlation function of the scattered light
intensity $g^{\left( 2\right) }\left( \tau \right) $ has not been
detailed and there also exist one assumption that the optical and
hydrodynamic radii of nano-particles are same.

\noindent For dilute homogeneous spherical particles, one method\cite{re15} is proposed to measure
the size distribution of nano-particles in dispersion accurately. The size 
distribution of nano-particles in dispersion is chosen to be a Gaussian distribution.
Using a non-linear least square fitting program, 
the mean static radius $\left\langle R_{s}\right\rangle $ and standard deviation $\sigma$
are measured accurately. Based on our research results, for the polystyrene spherical particles investigated, 
the size information obtained using the SLS technique and the commercial size information obtained
using Transmission Electron Microscopy (TEM) technique provided by the
supplier are consistent respectively. They all have a large difference with the corresponding hydrodynamic radii
respectively. Using the commercial particle size information, with one assumption, the calculated results and
measurements
of $g^{\left( 2\right)}\left( \tau \right)$ at all the scattering angles investigated are very well
consistent. This result also reveals that the static and hydrodynamic radii of
spherical nano-particles are different and can have a large difference. 
For the PNIPAM samples, the fitting results obtained using the non-linear least square fitting program are
very well consistent with the data of scattered light intensity and the residuals are random. 
Based on the static particle size distribution, with one assumption,
the calculated results and the measurements 
of $g^{\left( 2\right)}\left( \tau \right)$ at all the scattering angles investigated are also very well consistent. 
The results also show that the static and hydrodynamic radii of
spherical nano-particles are different and can have a much more large difference.

\noindent In order to investigate the characteristics of nano-particles in dispersion further, some researchers believe that 
there exist some relations between the physical quantities obtained using the
SLS and DLS techniques. A lot of people believe that the dimensionless shape
parameter  ${\left\langle{R_g}^2\right\rangle^{\frac{1}{2}}}/{R_{app,h}}$ can give a good description for the 
shapes of nano-particles. However based on our analyses, the dimensionless shape parameter
is ${\left\langle{R_g}^2\right\rangle^{\frac{1}{2}}}/{\left\langle{R_{s}}\right\rangle
}$. Therefore without the knowledge about 
the relationship of mean static and apparent hydrodynamic radii, this dimensionless
shape parameter ${\left\langle{R_g}^2\right\rangle^{\frac{1}{2}}}/{R_{app,h}}$ does not make sense.

\section{THEORY}

For vertically incident polarized light, the average
scattered light intensity of a dilute non-interacting particle system in
unit volume can be calculated using the following equation for homogeneous spherical particles and when
Rayleigh-Gans-Debye(RGD) approximation is valid
\begin{equation}
\frac{I_{s}}{I_{inc}}=\frac{4\pi ^{2}\sin ^{2}\theta
_{1}n_{s}^{2}\left( \frac{dn}{dc}\right) _{c=0}^{2}c}{\lambda
^{4}r^{2}}\frac{4\pi \rho }{3} \frac{\int_{0}^{\infty
}R_{s}^{6}P\left( q,R_{s}\right) G\left( R_{s}\right)
dR_{s}}{\int_{0}^{\infty }R_{s}^{3}G\left( R_{s}\right) dR_{s}},
\label{mainfit}
\end{equation}
where $ I_{s}$ is the intensity of the scattered light that reaches the detector,
$I_{inc}$ is the incident light intensity,
$R_{s}$ is the static radius of a particle, $\ q=\frac{4\pi
}{\lambda }n_{s}\sin \frac{\theta }{2}$ is the scattering vector,
$\lambda $ is the wavelength of incident light in vacuo, $n_{s}$\
is the solvent refractive index, $ \theta $ is the scattering
angle, $\rho $ is the density of particles, $r$
is the distance between the scattering particle and the point of
intensity measurement, $c$ is the mass concentration of particles, 
$\theta _{1}$ is the angle between the polarization of the
incident electric field and the propagation direction of the
scattered field, 
$P\left( q,R_{s}\right) $ is the form factor of homogeneous
spherical particles

\begin{equation}
P\left( q,R_{s}\right) =\frac{9}{q^{6}R_{s}^{6}}\left( \sin \left(
qR_{s}\right) -qR_{s}\cos \left( qR_{s}\right) \right) ^{2}
\label{P(qr)}
\end{equation}
and $G\left( R_{s}\right) $ is the number distribution of
particles in dispersion and is chosen as a
Gaussian distribution

\begin{equation}
G\left( R_{s};\left\langle R_{s}\right\rangle ,\sigma \right)
=\frac{1}{ \sigma \sqrt{2\pi }}\exp \left( -\frac{1}{2}\left(
\frac{R_{s}-\left\langle R_{s}\right\rangle }{\sigma }\right)
^{2}\right) ,
\end{equation}
where $\left\langle R_{s}\right\rangle $ is the mean static radius
and $\sigma $ is the standard deviation.

\noindent Based on the particle size information obtained using the SLS technique, 
for dilute homogeneous spherical particles, $g^{\left(1\right)}(\tau )$ is

\begin{equation}
g^{\left( 1\right) }\left( \tau \right) =\frac{\int R_{s}^{6}
P\left( q,R_{s}\right)G\left( R_{s}\right) \exp \left( -q^{2}D\tau
\right) dR_{s}}{\int R_{s}^{6}P\left( q,R_{s}\right) G\left(
R_{s}\right) dR_{s}}, \label{Grhrs}
\end{equation}
where $D$ is the diffusion coefficient.

\noindent Here the Stokes-Einstein relation\cite{re16} is still considered to be 
true for nano-particles in dispersion,

\begin{equation}
D=\frac{k_{B}T}{6\pi \eta _{0}R_{h}},
\end{equation}
where $\eta _{0}$, $k_{B}$ and $T$ are the viscosity of solvent,
Boltzmann's constant and absolute temperature respectively, then the
hydrodynamic radius $R_{h}$ can be obtained.

\noindent The relation between the static and
hydrodynamic radii in this work is assumed to be
\begin{equation}
R_{h}=kR_{s},  \label{RsRh}
\end{equation}
where $k$ is a constant. Based on the Siegert relation between
$g^{\left( 2\right) }\left( \tau \right) $ and $g^{\left( 1\right)
}\left( \tau \right)$ \cite{re7}

\begin{equation}
g^{\left( 2\right) }\left( \tau \right) =1+\beta \left( g^{\left(
1\right) }\right) ^{2},  \label{G1G2}
\end{equation}
the function between the SLS and DLS techniques is built and the values of $
g^{\left( 2\right) }\left( \tau \right) $ can be expected based on
the particle size information measured using the SLS technique.

\section{TRADITIONAL ANALYSIS}

In general, Static Light Scattering is a technique that can used to measure
the root mean square radius or radius of gyration $\left\langle R_{g}^{2}\right\rangle ^{1/2}$
in dispersion by measuring the scattered light intensity at many scattering angles. The values can 
be obtained using the following Zimm plot analysis.
\begin{equation}
Kc/R_{vv}= 1/M(1+q^2 \left\langle R_{g}^{2}\right\rangle/3)
\end{equation}
where $R_{vv}$ is the Rayleigh ratio $r^2I_s/I_{inc}$, $\left\langle R_{g}^{2}\right\rangle ^{1/2}$
is the radius of gyration.
 
\noindent Traditionally for non-interacting mono-disperse nano-particles in dispersion,
 $g^{\left( 1\right) }\left( \tau \right) $ can be written as

\begin{equation}
g^{\left( 1\right) }\left( \tau \right) =exp{\left(- \Gamma \tau \right)}
\end{equation}
where $\Gamma = q^2D$ is the decay rate, $D$ represents the macromolecular translational
diffusion coefficient of the particles.

\noindent For a polydisperse system, $g^{\left( 1\right) }\left( \tau \right) $ can be written as

\begin{equation}
g^{\left( 1\right) }\left( \tau \right) = \int{G{\left(\Gamma \right)}exp{\left(- \Gamma \tau \right)}d\Gamma}
\end{equation}
where $G{\left(\Gamma \right)}$ is the normalized distribution of the decay rates.

\noindent  The size distribution can be obtained using the method of moment analysis. The mean decay rate $\overline \Gamma$
and the moments of the distribution $\mu_2$ are defined as 

\begin{equation}
\overline \Gamma = \int{\Gamma G{\left(\Gamma \right)}d\Gamma}
\end{equation}
\begin{equation}
\mu_2 =  \int{{\left(\Gamma -\overline \Gamma\right)}^2 G{\left(\Gamma \right)}d\Gamma}
\end{equation}

\noindent Based on the Stokes-Einstein relation,  an apparent hydrodynamic radius $R_{app,h}$ can be defined as

\begin{equation}
D=\frac{k_{B}T}{6\pi \eta _{0}R_{app,h}}=\overline \Gamma /q^2 ,
\end{equation}

\noindent The polydispersity index also can be defined as

\begin{equation}
PD.I = \frac{\mu_2}{{\overline \Gamma}^2}.
\end{equation}

\noindent However the relation between the polydispersity index, apparent hydrodynamic radius $R_{app,h}$
and the particle size distribution does not be detailed.

\noindent The dimensionless shape parameter $\rho$ is defined as
\begin{equation}
\rho = {\left\langle{R_g}^2\right\rangle^{\frac{1}{2}}}/{R_{app,h}}
\end{equation}
are used to infer particle shapes. However based on the definition of 
the root mean-square radius of gyration $\left\langle R_{g}^{2}\right\rangle ^{1/2}$,
the definition $\rho$ is
\begin{equation}
\rho = {\left\langle{R_g}^2\right\rangle^{\frac{1}{2}}}/{\left\langle R_{s}\right\rangle }
\end{equation}

\section{RESULTS AND DISCUSSION}

\subsection{Standard polystyrene latex samples}

The commercial particle size information was measured by using the TEM technique provided by the supplier. 
Since there exists a large difference between the refractive indices of the
polystyrene latex (1.591 at wavelength 590 nm and 20 $^\mathrm
o$C) and water (1.332), i.e., the ``phase shift'' $\frac{4\pi
}{\lambda }R|m-1|$ \cite{re6, re14} of  Latex-1
and Latex-2 are 0.13 and 0.21 respectively. They do not exactly satisfy the rough
criterion that a RGD approximation\cite{re6} is valid. Therefore 
the mono-disperse model $G\left( R_{s}\right) =\delta \left(
R_{s}-\left\langle R_{s}\right\rangle \right) $ was used to
obtain the values of $\left\langle R_{s}\right\rangle $.
 All the results measured using the
SLS technique and TEM are listed in Table \ref{table2}. The fitting results and measured 
data are shown in Figs. \ref{figPolyfitcal}a and \ref{figPolyfitcal}b, respectively.
The results reveal that the values measured using the SLS and TEM techniques are very well
consistent.

\begin{table}[ht]
\begin{center}
\begin{tabular}{|c|c|c|}
\hline $\left\langle R\right\rangle_{comm}$ (nm) & $\sigma_{comm}$
(nm) & $\left\langle R_{s}\right\rangle$ (nm)\\
\hline 33.5(Latex-1) & 2.5 & 33.3$\pm $0.2 \\
\hline 55(Latex-2) & 2.5 & 56.77$\pm $0.04 \\
\hline
\end{tabular}
\caption[]{Size information measured using the TEM and SLS
techniques.}\label{table2}
\end{center}
\end{table}

\begin{center}
  $\begin{array}{c@{\hspace{0in}}c}
     \includegraphics[width=0.35\textwidth,angle=-90]{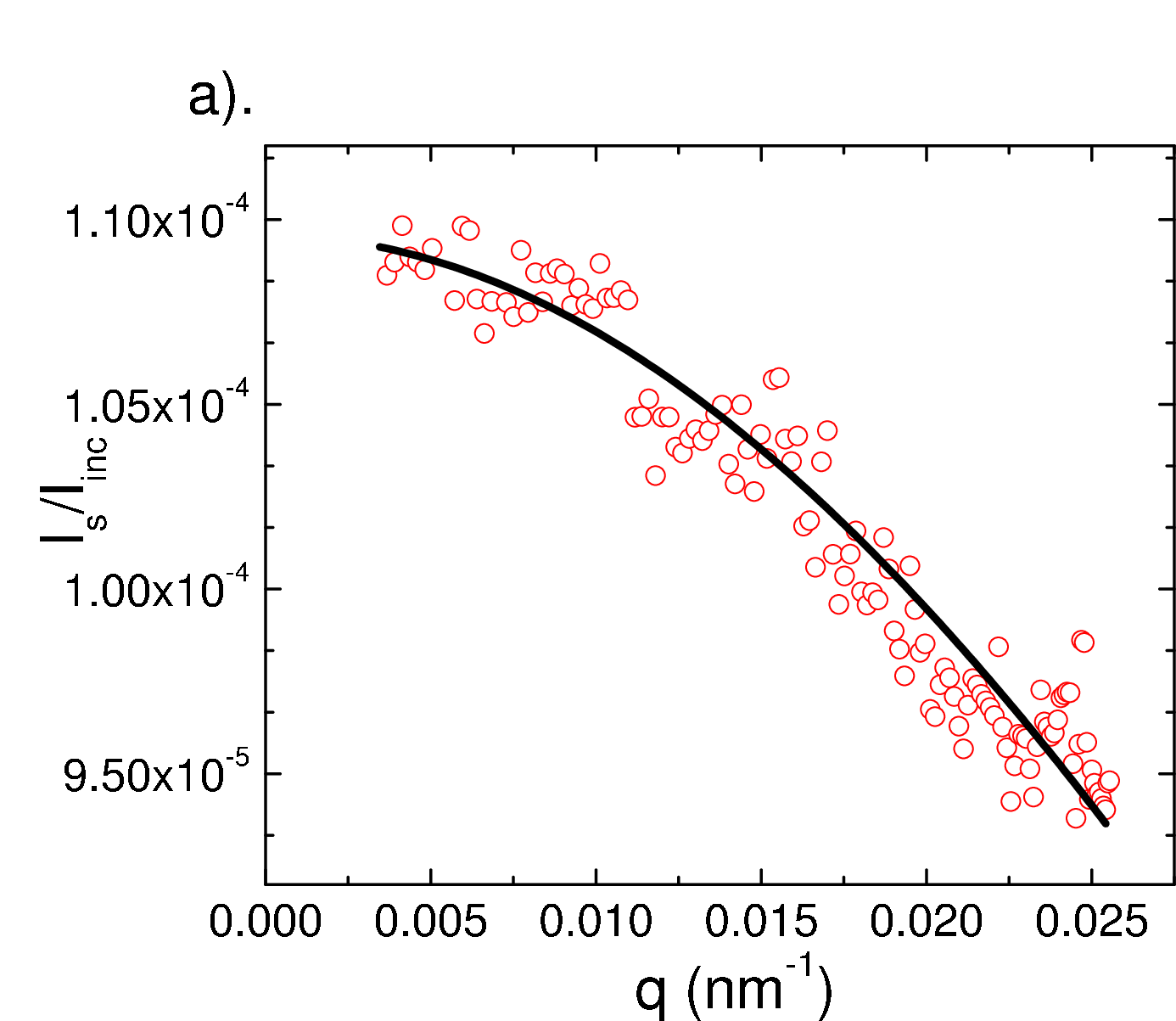} &
     \includegraphics[width=0.35\textwidth,angle=-90]{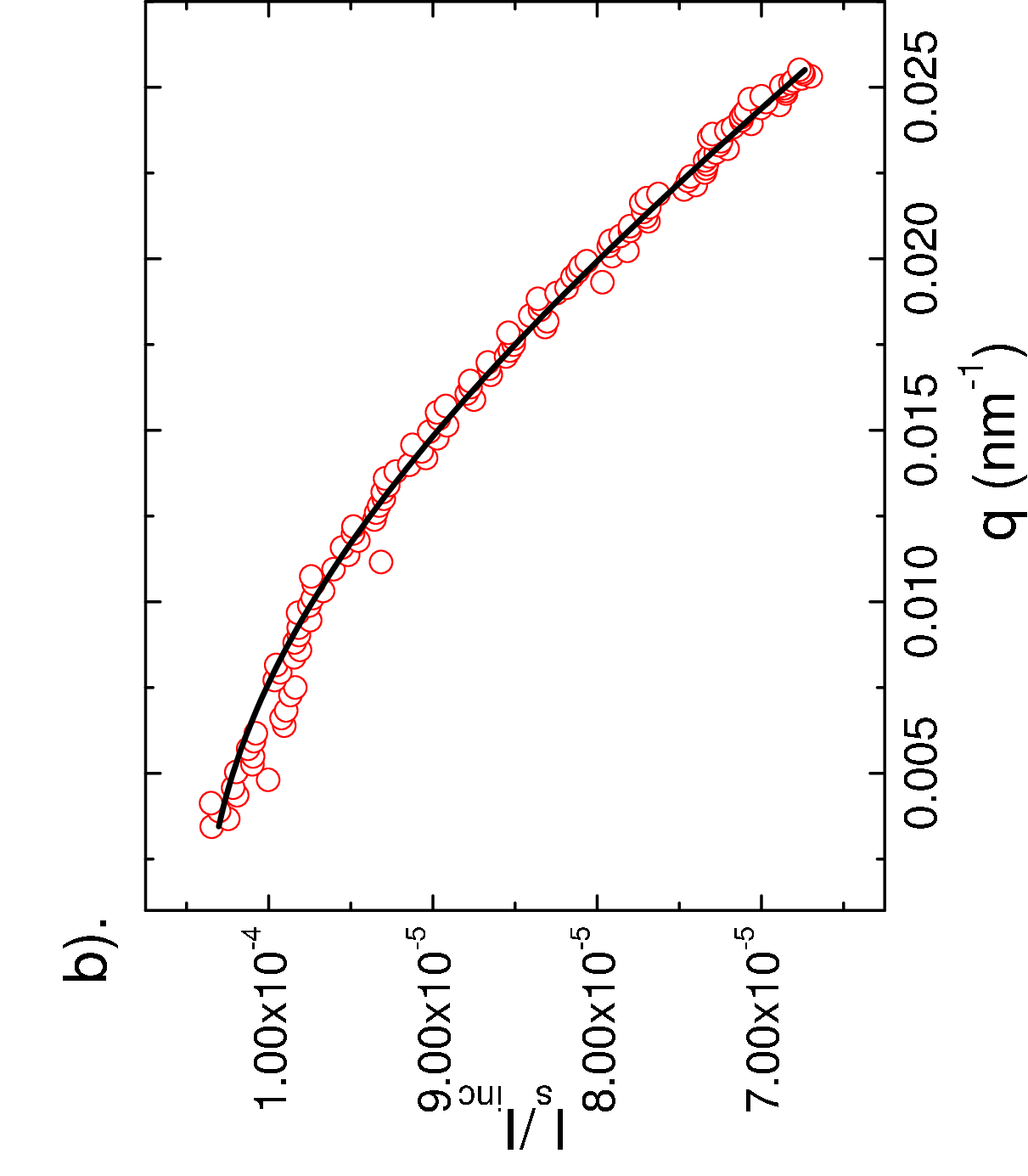} \\ [0.0cm]
    \end{array}$
   \end{center}\vspace{-0.5cm}
 \makeatletter\def\@captype{figure}\makeatother
\caption[] {a). The experimental data and fitting results of
$I_{s}/I_{inc}$ for Latex-1 and b). Latex-2. The circles 
show the experimental data and the line
represents the fitting results of $I_{s}/I_{inc}$.} \label{figPolyfitcal}

\noindent  Since the size information measured using the SLS and TEM techniques are consistent,
so the particle size information provided by the supplier are used to calculate the expected values 
of $g^{\left( 2\right) }\left( \tau \right) $. In these calculations,
the constant $k$ in Eq. \ref{RsRh} is set to be 1.1 for
Latex-1 and 1.2 for Latex-2. Figs.
\ref{figPolyDLScal}a and \ref{figPolyDLScal}b show 
all the experimental data and expected values of
$g^{\left( 2\right) }\left( \tau \right) $ at the scattering
angles 30$^\mathrm o$, 60$^\mathrm o$, 90$^\mathrm o$,
120$^\mathrm o$ and 150$^\mathrm o$ and a temperature of 298.45 K
for Latex-1, 298.17 K for Latex-2 respectively.
Figure \ref{figPolyDLScal} shows that the experimental data and expected values are
very well consistent.

\begin{center}
  $\begin{array}{c@{\hspace{0in}}c}
     \includegraphics[width=0.35\textwidth,angle=-90]{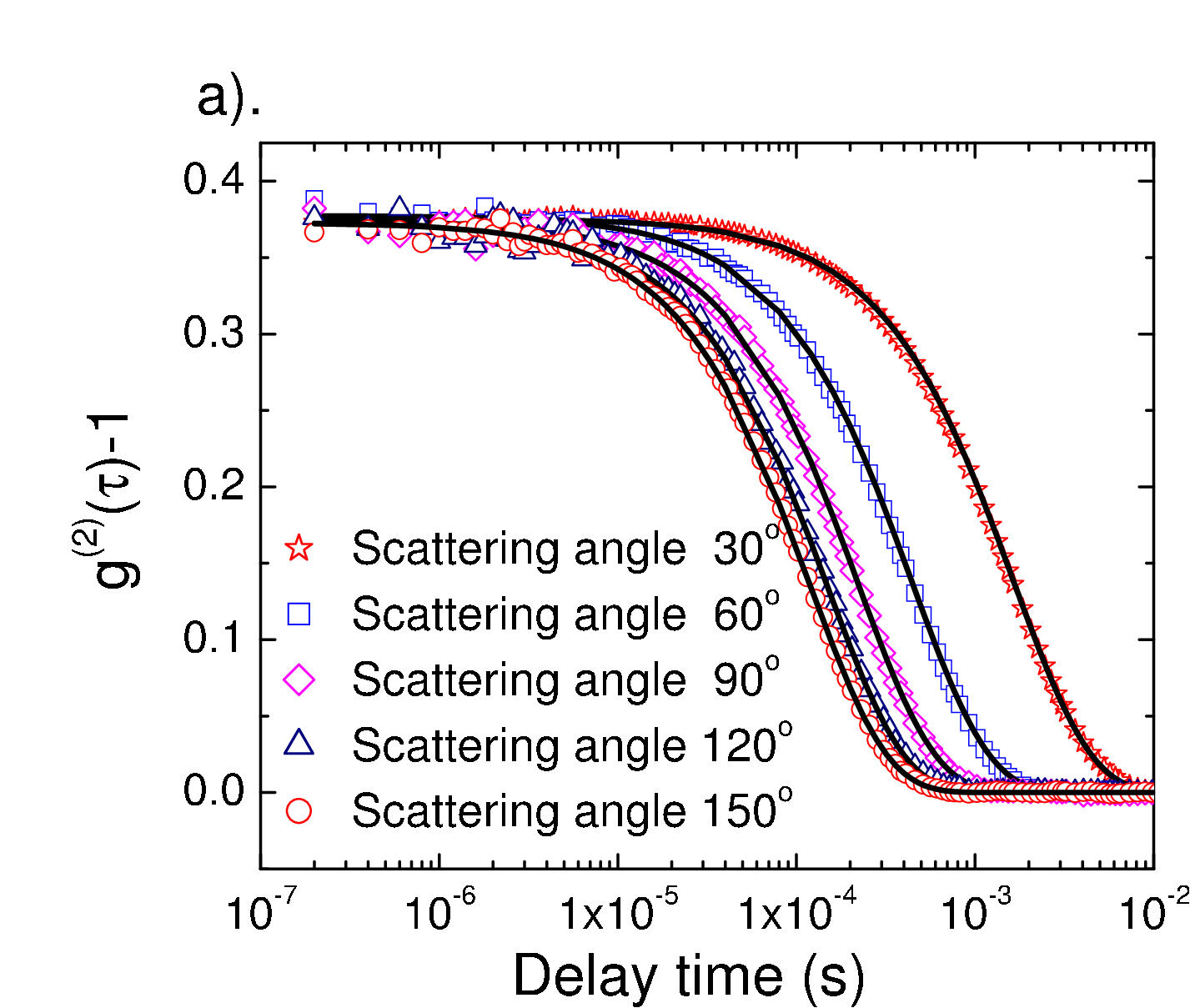} &
     \includegraphics[width=0.35\textwidth,angle=-90]{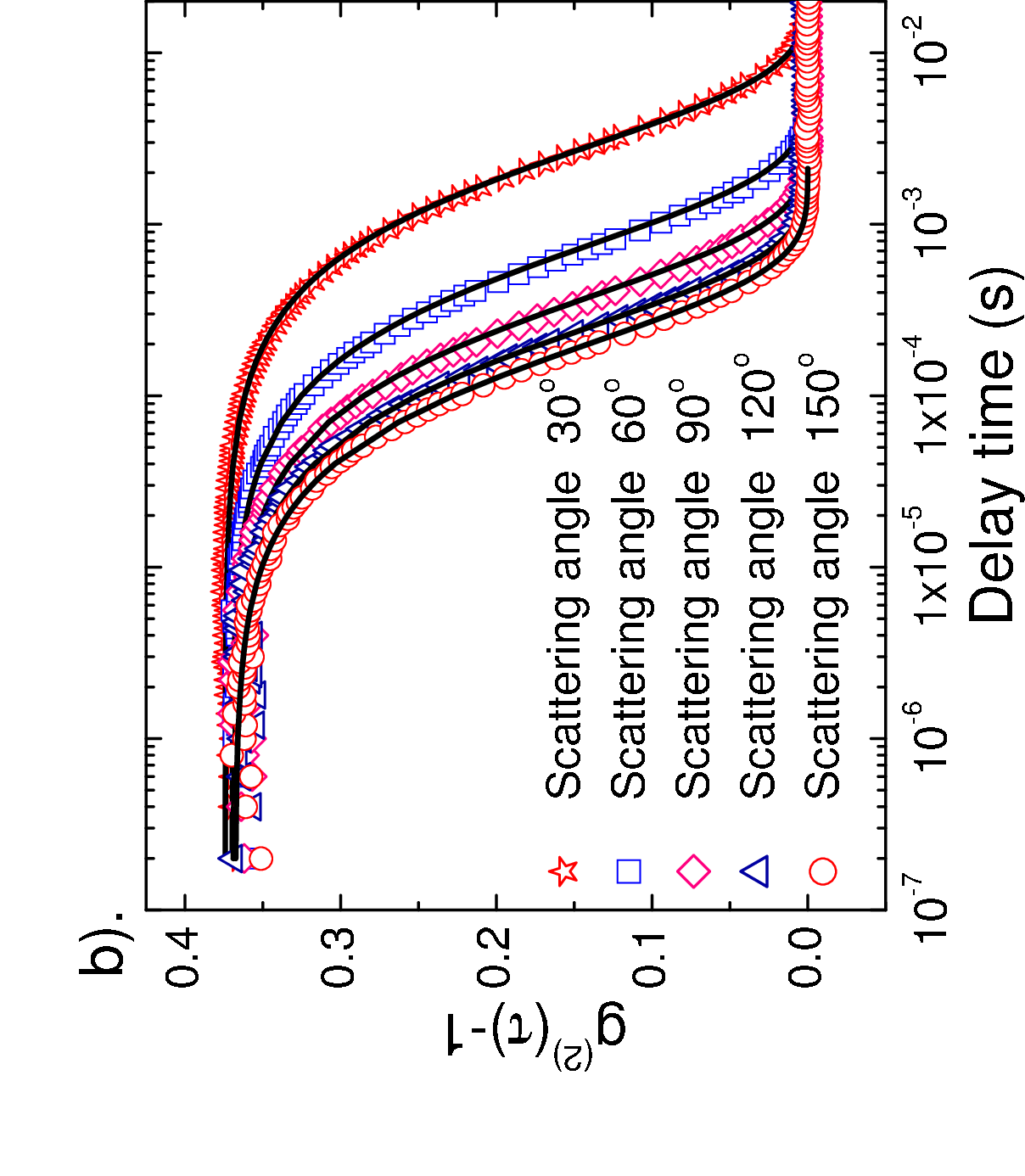} \\ [0.0cm]
    \end{array}$
   \end{center}\vspace{-0.5cm}
 \makeatletter\def\@captype{figure}\makeatother
  \caption{The experimental data and expected values of $g^{\left(
2\right) }\left( \tau \right) $. a). Latex-1 and b).
Latex-2. The symbols represent the experimental data and the lines show
the expected values calculated under $R_{h}=kR_{s}$.}
\label{figPolyDLScal}

\subsection{PNIPAM samples}

When the data of the PNIPAM-1 measured using the SLS technique
at a temperature of 302.33 K was fitted using Eq. \ref{mainfit},
the fitting results show that if a small scattering vector range is chosen, the
parameters are not well-determined.  The uncertainties in the parameters will decrease and
$\left\langle R_{s}\right\rangle $ and $\sigma $ stabilize if the scattering vector range
is increased. As the scattering vector range continues to increase, the values of $\left\langle
R_{s}\right\rangle $ and $\sigma $ begin to lose stability and $\chi ^{2}$
grows as shown in Table \ref{table3}. Therefore the stable
fitting results $\left\langle R_s\right\rangle = 254.3 \pm 0.1$ nm and
$\sigma = 21.5 \pm 0.3$ nm obtained in the scattering vector range
between 0.00345 and 0.01517 nm$^{-1}$ are chosen as the particle size
information measured using the SLS technique. All the experimental data,
stable fitting results and residuals in the scattering vector range between 0.00345 and
0.01517 nm$^{-1}$ are shown in Fig. \ref{figPNI-302fit}. The figure reveals that 
the fitting results and measured data are very well consistent and the residuals are random.

\begin{center}
\begin{tabular}{|c|c|c|c|}
\hline $q$ ($10^{-3}$ nm$^{-1}$) & $\left\langle
R_{s}\right\rangle$ (nm) & $ \sigma$ (nm) & $\chi ^{2}$ \\
\hline 3.45 to 9.05 & 260.09$\pm $9.81 & 12.66$\pm $19.81 & 1.64 \\
\hline 3.45 to 11.18 & 260.30$\pm $1.49 & 12.30$\pm $3.37 & 1.65 \\
\hline 3.45 to 13.23 & 253.45$\pm $0.69 & 22.80$\pm $0.94 & 2.26 \\
\hline 3.45 to 14.21 & 254.10$\pm $0.15 & 21.94$\pm $0.36 & 2.03 \\
\hline 3.45 to 15.17 & 254.34$\pm $0.12 & 21.47$\pm $0.33 & 2.15 \\
\hline 3.45 to 17.00 & 255.40$\pm $0.10 & 17.32$\pm $0.22 & 11.02 \\
\hline
\end{tabular}
 \makeatletter\def\@captype{table}\makeatother
\caption{The results obtained using Eq. \ref{mainfit} for
PNIPAM-1 at different scattering vector ranges and a temperature
of 302.33 K.}\label{table3}
\end{center}

\begin{center}
   \includegraphics[width=0.35\textwidth,angle=-90]{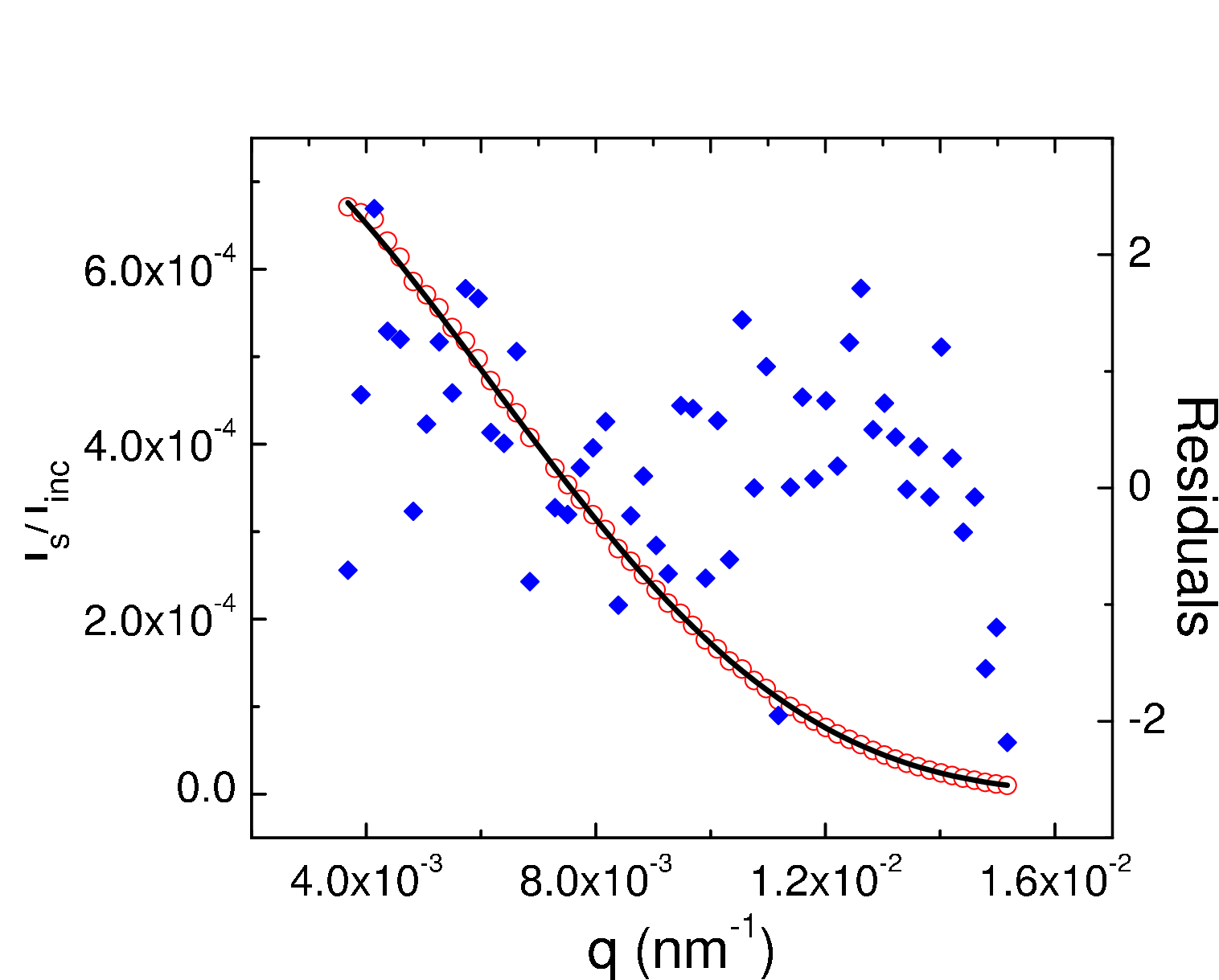}
      \makeatletter\def\@captype{figure}\makeatother
   \caption{The experimental data and stable fitting results for PNIPAM-1 at a temperature of 302.33 K.
   The circles show the experimental data, the line shows the fitting results and the diamonds show the
residuals: $\left( y_{i}-y_{fit}\right) /\sigma _{i}$.}
 \label{figPNI-302fit}
\end{center}

\noindent  Based on the particle size information measured using the SLS technique with the constant $k=1.21$ in Eq. \ref{RsRh}, 
the expected values of $g^{\left( 2\right) }\left( \tau \right) $ at the scattering
angles 30$^\mathrm o$, 50$^\mathrm o$ and 70$^\mathrm o$ are calculated. 
All the measured data and expected values are shown in Fig. \ref{figPNI-302cal}.
Figure \ref{figPNI-302cal} reveals that
the expected values and experimental data are very well consistent.

\begin{center}
  \includegraphics[width=0.35\textwidth,angle=-90]{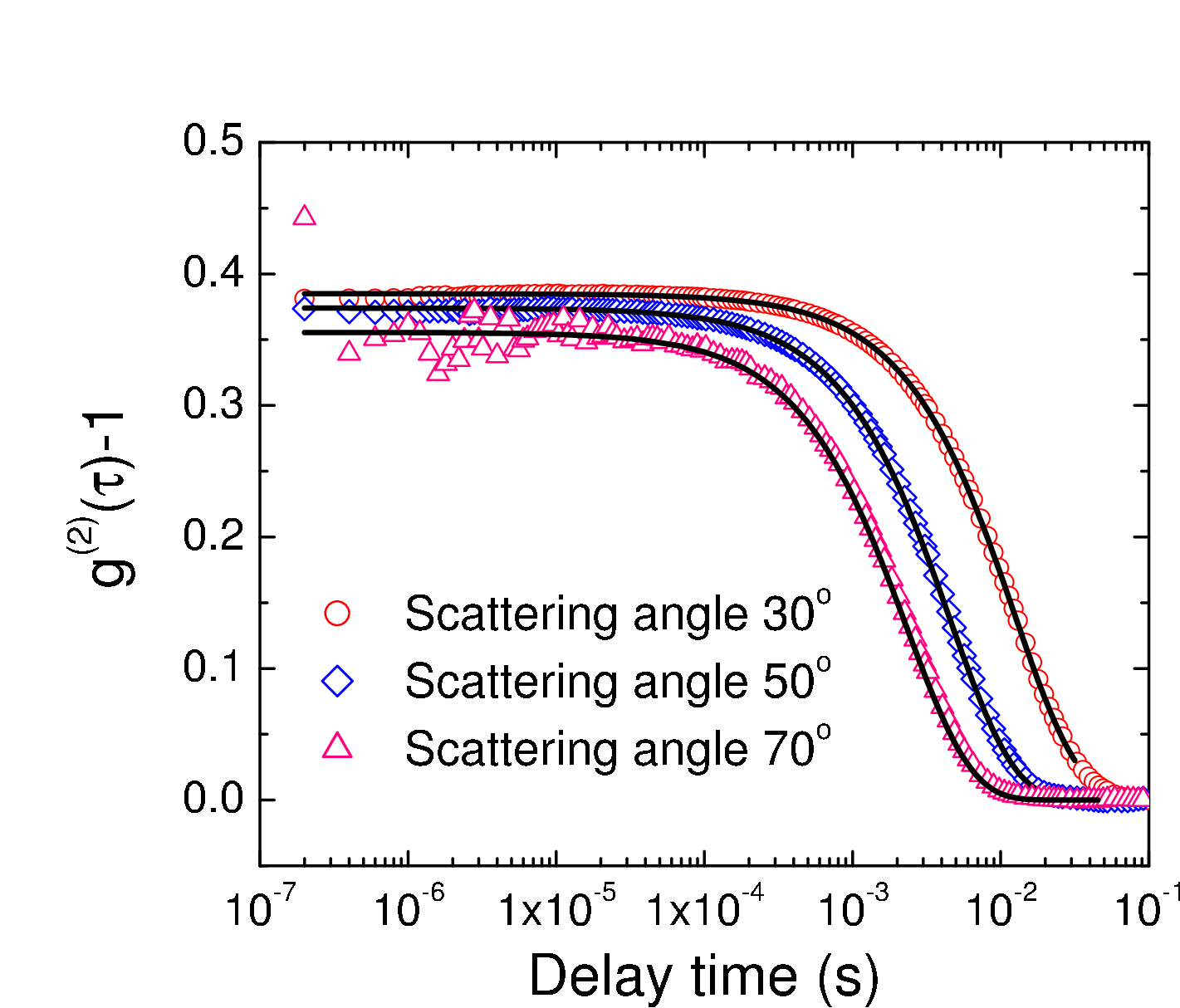}
   \makeatletter\def\@captype{figure}\makeatother
  \caption{The measured data and expected values of $g^{\left(
2\right) }\left( \tau \right)$ of PNIPAM-1 at a temperature of
302.33 K. The symbols represent the experimental data and the lines
show the expected values calculated under $R_{h}=1.21R_{s}$.}
\label{figPNI-302cal}
\end{center}

\noindent  When the PNIPAM samples in dispersion are at high temperatures, all the situations that fit using
Eq. \ref{mainfit} are the same as that of PNIPAM-1 at a temperature
of 302.33 K. The fitting values of $\left\langle R_{s}\right\rangle $ and
$\sigma $ are affected by a scattering vector range. The uncertainties in the parameters will decrease and
$\left\langle R_{s}\right\rangle $ and $\sigma $ stabilize if the scattering vector range
is increased. For PNIPAM-5 at a temperature of
312.66 K,  the stable fitting
results $\left\langle R_s\right\rangle = 139.3 \pm 0.3$ nm and
$\sigma = 12.4 \pm 0.6$ nm obtained in the scattering vector range
between 0.00345 and 0.02555 nm$^{-1}$ are chosen as the particle size information measured using the SLS
technique. The stable fitting results and
residuals are shown in Figure \ref{figPNIhT}a. With the constant $k=1.1$ in Eq. \ref{RsRh}, the expected values of 
$g^{\left( 2\right) }\left( \tau \right) $ at the scattering
angles 30$^\mathrm o$, 50$^\mathrm o$, 70$^\mathrm o$ and 100$^\mathrm o$ are calculated. 
All the experimental and expected values are shown in Fig. \ref{figPNIhT}b. The expected 
values and measured data of $g^{\left( 2\right) }\left( \tau \right) $ also are very well 
consistent.

\begin{center}
  $\begin{array}{c@{\hspace{0in}}c}
     \includegraphics[width=0.35\textwidth,angle=-90]{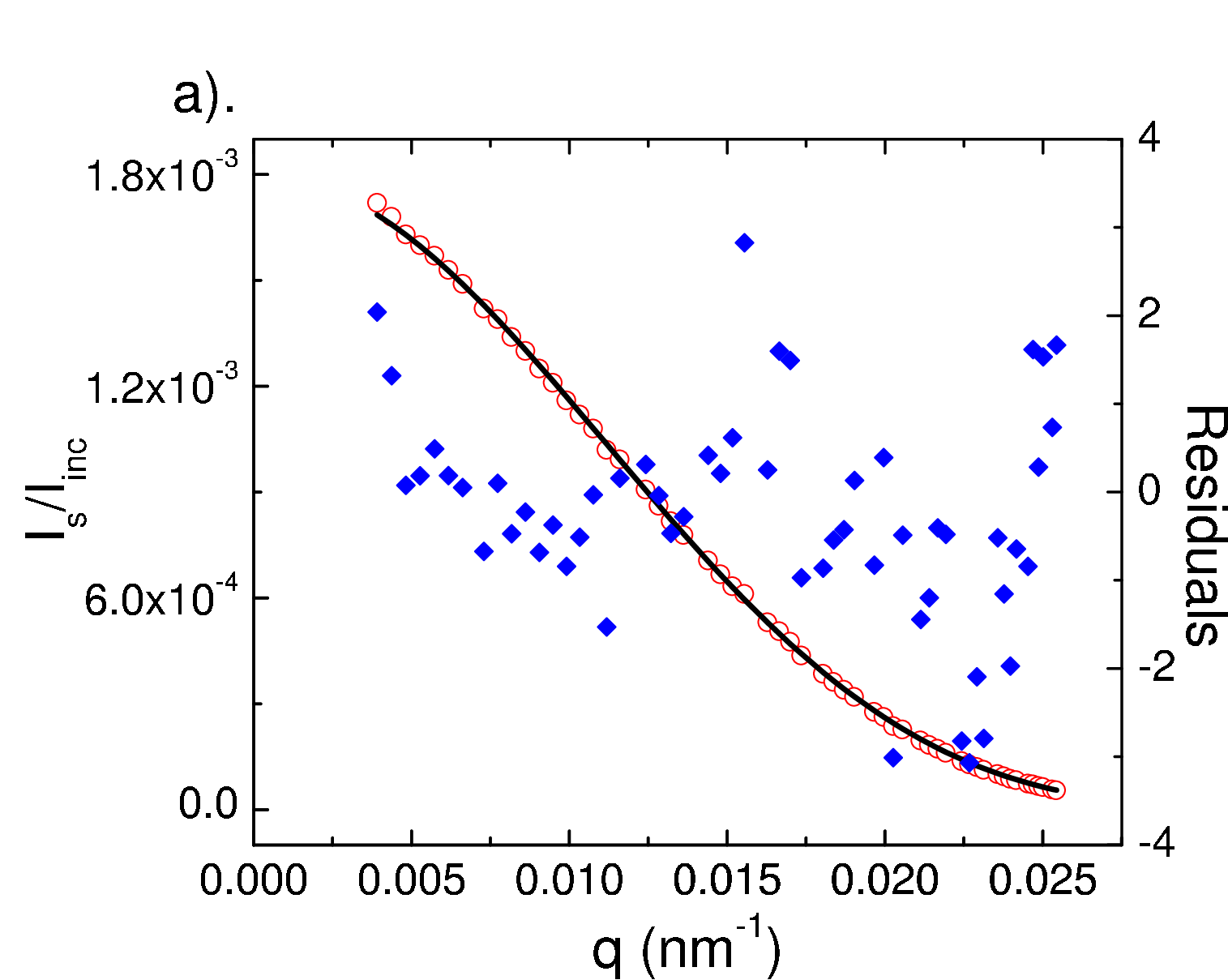} &
     \includegraphics[width=0.35\textwidth,angle=-90]{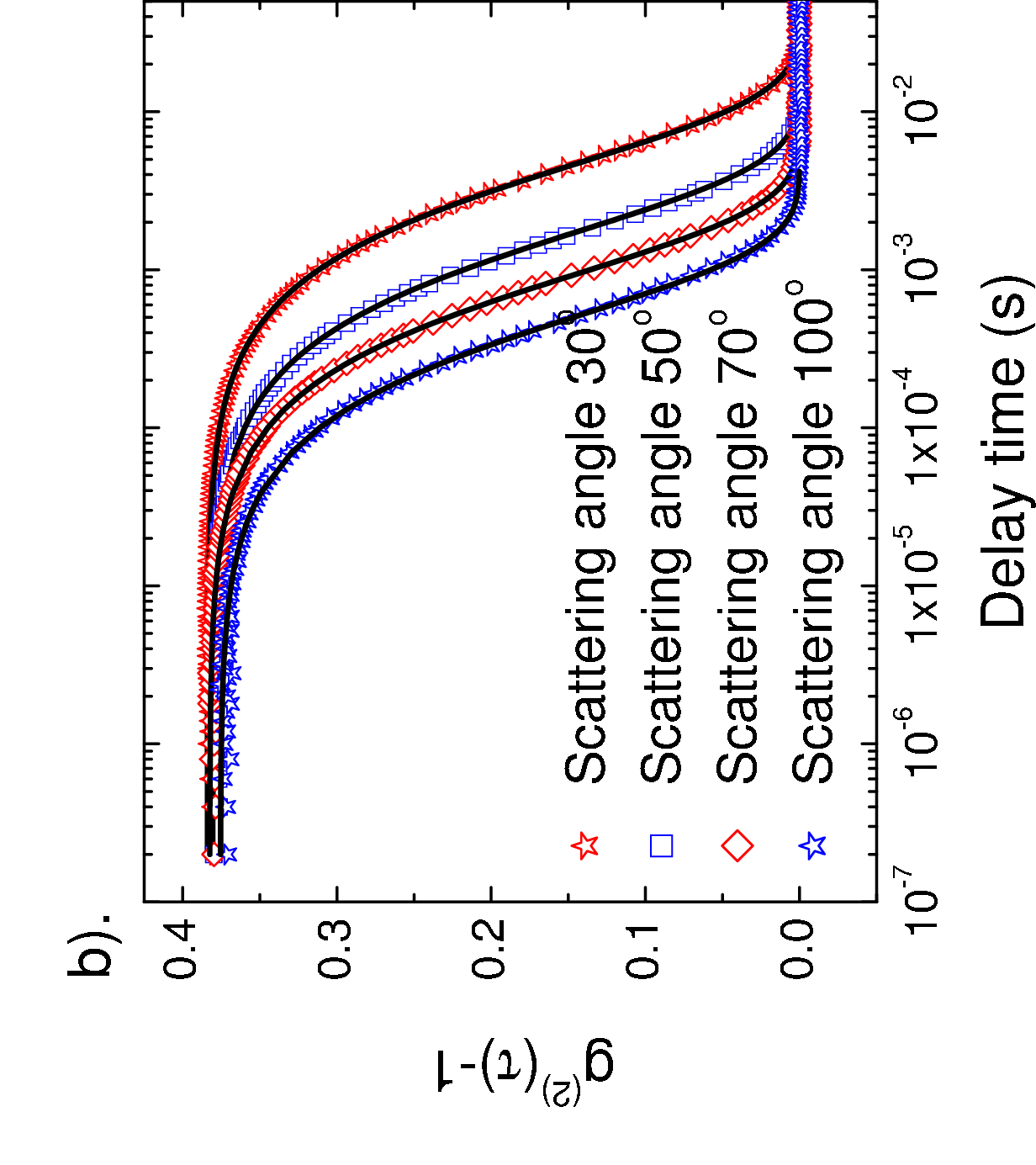} \\ [0.0cm]
    \end{array}$
   \end{center}\vspace{-0.5cm}
 \makeatletter\def\@captype{figure}\makeatother
  \caption{The static and dynamic results of PNIPAM-5 at a temperature 312.66 K. 
  a). The measured data and stable fitting results. The circles represent the experimental data, the line shows the fitting results and the diamonds represent the
residuals: $\left( y_{i}-y_{fit}\right) /\sigma _{i}$. b). The
measured data and expected values of $g^{\left( 2\right)
}\left( \tau \right)$. The symbols represent the measured data and
the lines show the expected values calculated under
$R_{h}=1.1R_{s}$.} \label{figPNIhT}

\noindent  For all other samples investigated, the same situation has been obtained.
The fitting results of $\left\langle R_{s}\right\rangle
$ and $\sigma $ are affected by a scattering vector range.
If a small scattering vector range is used, the parameters are
not well-determined. The uncertainties in the parameters will decrease and $\left\langle
R_{s}\right\rangle $ and $\sigma $ stabilize if the scattering vector range is increased. 
The expected values of $g^{\left( 2\right) }\left( \tau \right) $
 calculated based on the particle size distribution
measured using the SLS technique or commercial results provided by the
supplier and measured data are very well consistent.
All the results reveal that the static radius of nano-particles in dispersion can have a very large
difference with hydrodynamic radius.

\noindent  For the DLS technique, two important parameters that people try to measure
accurately from the DLS data are an apparent hydrodynamic radius and polydispersity
index. People also believe that the
values of polydispersity index represent the size distribution of nano-particles. 
The small values mean the size distribution is narrow or mono-disperse 
then the apparent hydrodynamic radius will be equal to the mean hydrodynamic radius.
Since the apparent hydrodynamic radius and polydispersity index of nano-particles are
measured from the non-exponentility of
the correlation function of temporal fluctuation in the scattered
light at a given scattering angle, therefore the effects of nano-particle size distribution 
and scattering angles on the correlation function of temporal fluctuation in the scattered light
were investigated further.
For simplicity, let $k=1$ that means the values of the static and hydrodynamic
radii of homogeneous spherical particles are same.

\noindent  Using the Cumulant method at a given scatter angle $q$ as $\tau\rightarrow0 $, the equations of 
the apparent hydrodynamic radius ${R_{app,h}}$ and polydispersity
index $PD.I$ are 

\begin{equation}
{R_{app,h}}=\frac{\int_{0}^{\infty }R_{s}^{6}P\left( q,R_{s}\right)
G\left( R_{s}\right) dR_{s}}{\int_{0}^{\infty }R_{s}^{5}P\left(
q,R_{s}\right) G\left( R_{s}\right) dR_{s}} \label{Rh}
\end{equation}

\noindent and

\begin{equation}
PD.I=\frac{\int_{0}^{\infty }R_{s}^{4}P\left( q,R_{s}\right)
G\left( R_{s}\right) dR_{s}\int_{0}^{\infty }R_{s}^{6}P\left(
q,R_{s}\right) G\left( R_{s}\right) dR_{s}}{\left(
\int_{0}^{\infty }R_{s}^{5}P\left( q,R_{s}\right) G\left(
R_{s}\right) dR_{s}\right) ^{2}}-1 . \label{Dzindex}
\end{equation}

\noindent  The expected values of ${R_{app,h}}$ and $PD.I$ were calculated for the mean hydrodynamic
radius 5 nm and different standard deviations at scattering
angles 0$^\mathrm o$, 30$^\mathrm o$, 60$^\mathrm o$, 90$^\mathrm
o$ and 120$^\mathrm o$. All the results are shown in Figs. \ref{figRhPDS}a and
\ref{figRhPDS}b, respectively. During the process, the
wavelength of the incident light in vacuo $\lambda $ is set to
632.8 nm and solvent refractive index $n_{s}$ is 1.332.

\begin{center}
   $\begin{array}{c@{\hspace{0in}}c}
     \includegraphics[width=0.35\textwidth,angle=-90]{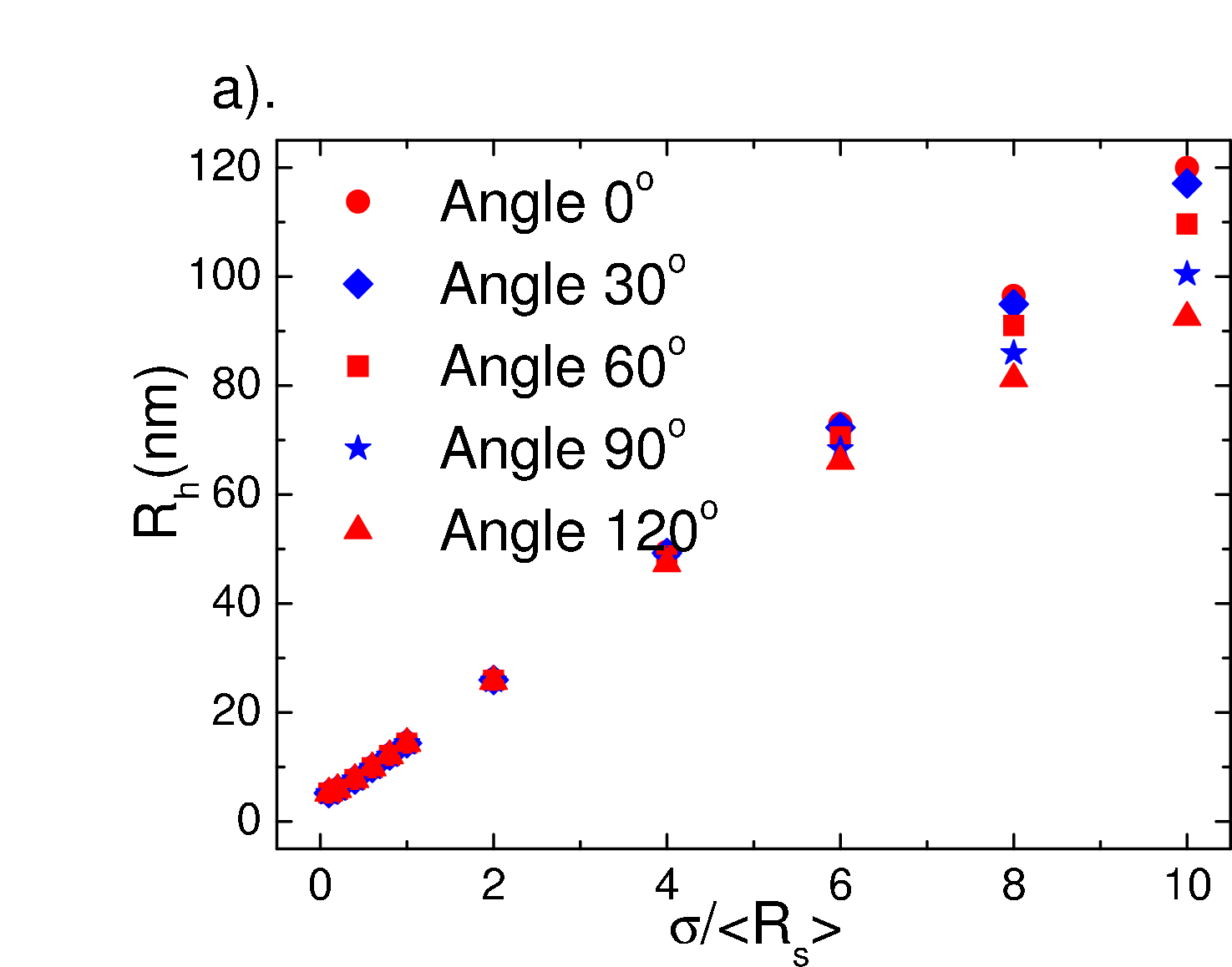} &
     \includegraphics[width=0.35\textwidth,angle=-90]{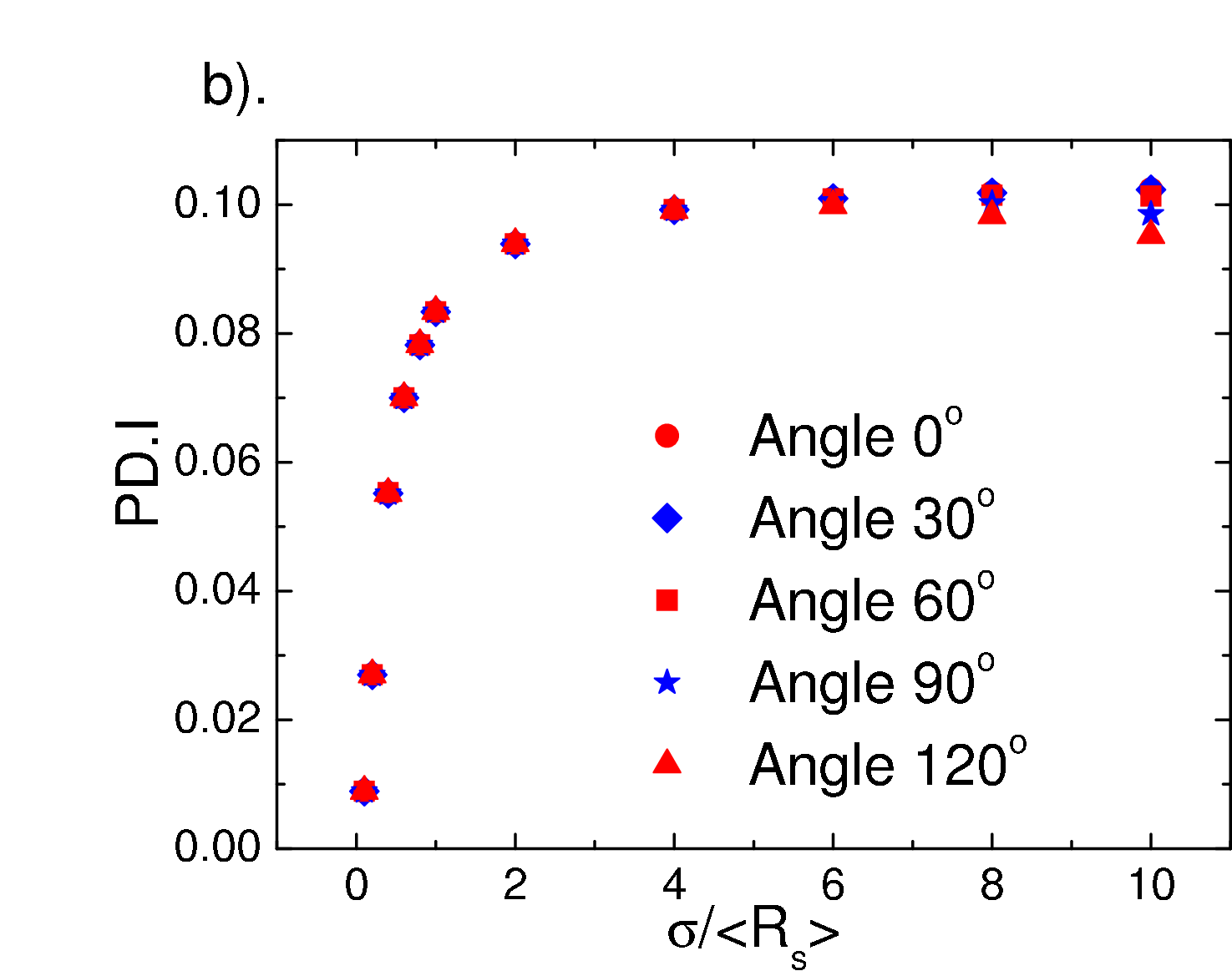} \\ [0.0cm]
   \end{array}$
\end{center}\vspace{-0.5cm}
 \makeatletter\def\@captype{figure}\makeatother
\caption[] {a). The apparent hydrodynamic radius and b).
polydispersity index as a function of relative standard deviation
at different scattering angles with
mean hydrodynamic radius 5 nm.} \label{figRhPDS}

\noindent  Figure \ref{figRhPDS}a reveals that the effects of the relative standard deviation
on the apparent hydrodynamic radius is huge.
 Figure \ref{figRhPDS}b shows that polydispersity index always is small and 
increase to about 0.1 as the values of relative standard deviation change to very large.

\noindent  Next the situation of large particles also was investigated.
The mean hydrodynamic radius is 100nm. The expected values of ${R_{app,h}}$ and $PD.I$
were calculated as a function of relative standard deviation at scattering angles 0$^\mathrm o$,
30$^\mathrm o$, 60$^\mathrm o$, 90$^\mathrm o$ and 120$^\mathrm o$.
The results are shown in Figs. \ref{figRhPDL}a and \ref{figRhPDL}b,
respectively.  Figure \ref{figRhPDL}a and  \ref{figRhPDS}b reveal that the effects of the relative standard
deviation and scattering angles on the apparent hydrodynamic radius and polydispersity index are huge and complex.

\noindent Based on the discussion above, even if the value of polydispersity index is small and the values of apparent hydrodynamic radii measured  are consistent at different scattering angles for an unknown sample, the conclusion that 
the particle size distribution is narrow or mono-disperse still cannot be obtained.

\begin{center}
   $\begin{array}{c@{\hspace{0in}}c}
     \includegraphics[width=0.35\textwidth,angle=-90]{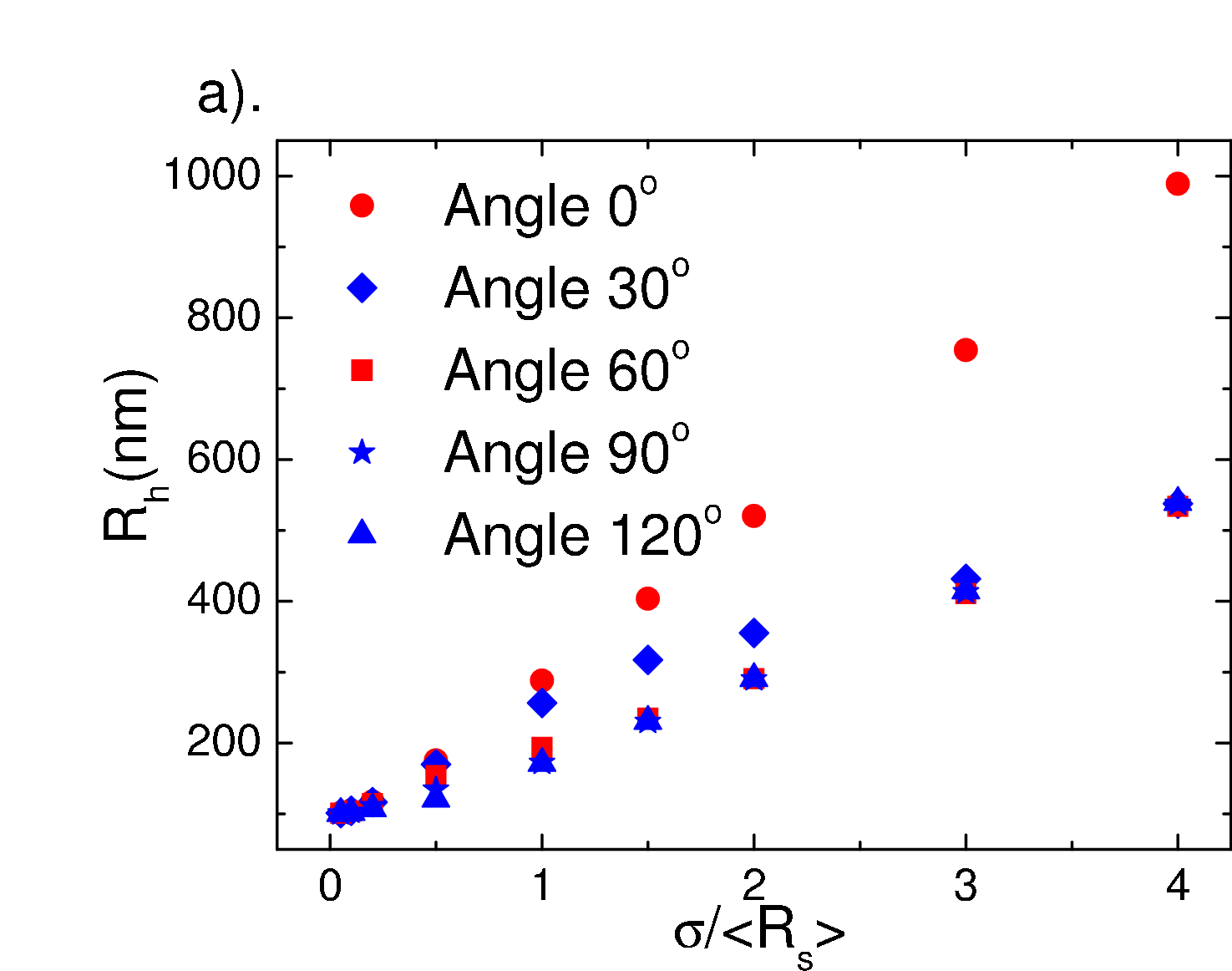} &
     \includegraphics[width=0.35\textwidth,angle=-90]{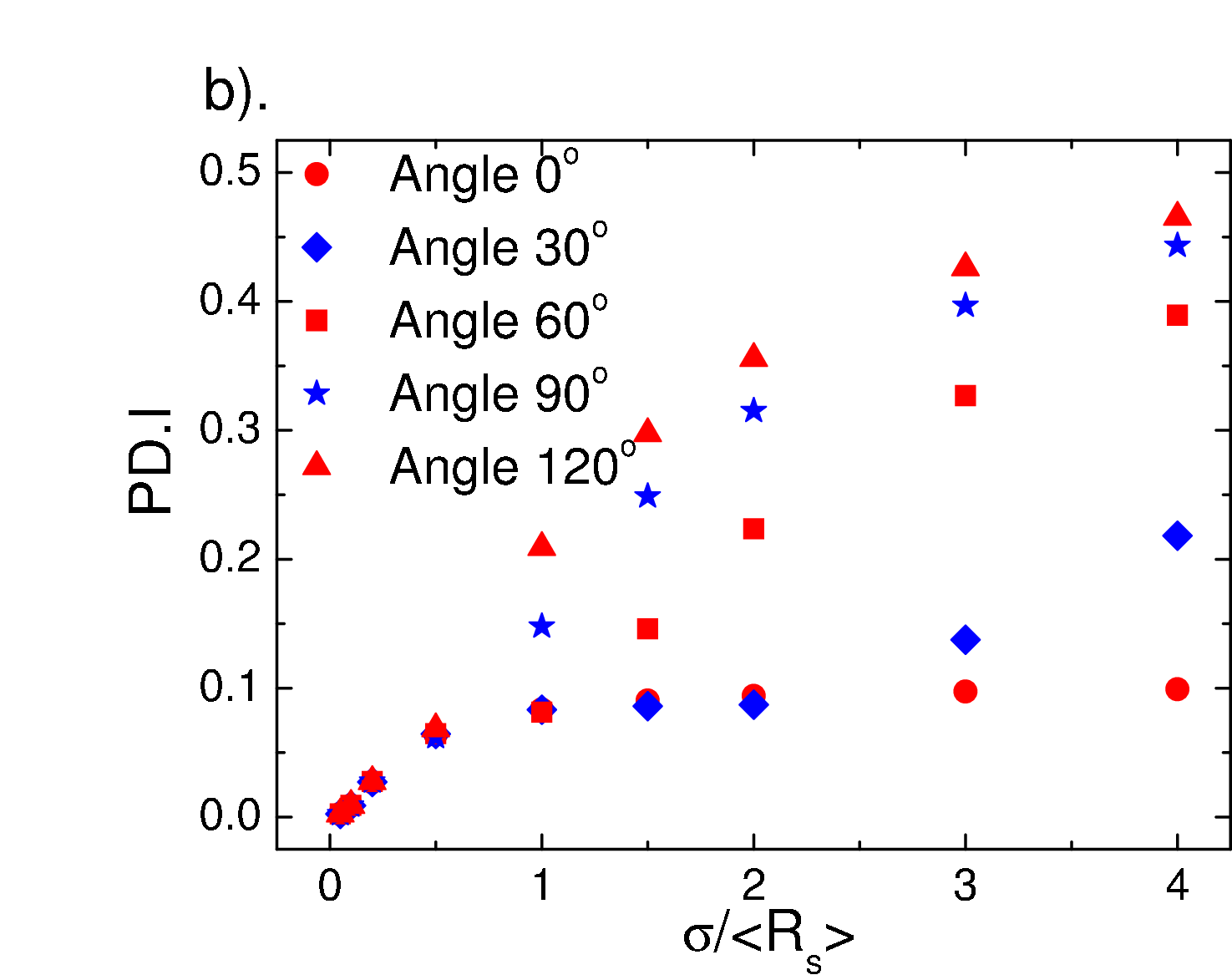} \\ [0.0cm]
   \end{array}$
\end{center}\vspace{-0.5cm}
 \makeatletter\def\@captype{figure}\makeatother
\caption[] {a). The apparent hydrodynamic radius and b).
polydispersity index as a function of relative standard deviation
at scattering angles 0$^\mathrm o$, 30$^\mathrm o$, 60$^\mathrm
o$, 90$^\mathrm o$ and 120$^\mathrm o$ for the nanoparticles with
mean hydrodynamic radius 100 nm.} \label{figRhPDL}

\noindent  For  the polystyrene latex samples investigated, the results obtained using the 
TEM, SLS and DLS techniques are listed in Table \ref{table6}. 

\begin{center}
\begin{tabular}{|c|c|c|c|}
\hline $R_{TEM}$(nm) & $\left\langle R_{s}\right\rangle$(nm) & $R_{app,h}$
(nm) & $R_{app,h}/\left\langle R_{s}\right\rangle$ \\
\hline 33.5 & 33.3$\pm $0.2 & 37.27$\pm $0.09 & 1.119$\pm$0.007 \\
\hline 55 & 56.77$\pm $0.04 & 64.5$\pm $0.6 & 1.14$\pm$0.01 \\
\hline 90 & 92.05$\pm $0.04 & 103.1$\pm $0.4 & 1.120$\pm$0.004 \\
\hline
\end{tabular}
 \makeatletter\def\@captype{table}\makeatother
\caption{ Commercial $R_{TEM}$, $\left\langle R_{s}\right\rangle$ and
apparent hydrodynamic radii.}
\label{table6}
\end{center}

\noindent  The results above reveal that the sizes obtained using the SLS and TEM techniques
are very well consistent. Even for very narrow samples, 
the value of the apparent hydrodynamic radius obtained under the same conditions
as the static radius is larger than that of the static radius by about $12\%$.

\noindent  Meanwhile the values of the root mean square radius of gyration 
${\left\langle {R_g}^2\right\rangle}^{1/2} $
also can be calculated using the commercial size distribution.  The expected 
and measured values obtained from the Zimm plot analysis are listed 
in Table \ref{table7}. All the results reveal that particle size information
 measured using the SLS and TEM are consistent.

\begin{center}
\begin{tabular}{|c|c|c|}
\hline Sample & ${\left\langle {R_g}^2\right\rangle}^{1/2}_{cal} $ & ${\left\langle 
{R_g}^2\right\rangle}^{1/2}_{Zimm} $\\
\hline 33.5(nm) & 26.9  & 26.9$\pm $0.5 \\
\hline 55 (nm)  & 43.24 & 46.8$\pm $0.3  \\
\hline 90 (nm) & 70.1   & 69.0$\pm $2.0 \\
\hline
\end{tabular}
 \makeatletter\def\@captype{table}\makeatother
\caption{Values of ${\left\langle {R_g}^2\right\rangle}^{1/2}_{cal} $ and
 ${\left\langle {R_g}^2\right\rangle}^{1/2}_{Zimm} $.}
\label{table7}
\end{center}

\noindent For the PNIPAM samples investigated, the dimensionless parameters of 
$\rho$, ${\left\langle {R_g}^2\right\rangle}^{1/2}_{Zimm} /\left\langle R_{s}\right\rangle$ and 
${\left\langle {R_g}^2\right\rangle}^{1/2}_{cal}/\left\langle R_{s}\right\rangle $
also have been measured or calculated respectively. All the results are shown in Table \ref{table8}.
All the results also reveal that the particle size information obtained using the SLS technique is
more accurate to represent the particle information in dispersion. The dimensionless
parameter $\rho$ cannot give a good description for the shapes of nano-particles in dispersion.

\begin{center}
\begin{tabular}{|c|c|c|c|}
\hline Sample (Temperature) & $\rho$  & ${\left\langle {R_g}^2\right\rangle}^{1/2}_{Zimm} /\left\langle R_{s}\right\rangle$ & ${\left\langle {R_g}^2\right\rangle}^{1/2}_{cal}/\left\langle R_{s}\right\rangle $ \\
\hline PNIPAM-5($40^o$C) & 0.73$\pm $0.02 & 0.83$\pm $0.03 & 0.813$\pm$0.003 \\
\hline PNIPAM-2($40^o$C) & 0.69$\pm $0.03 & 0.84$\pm $0.04 & 0.82$\pm$0.01 \\
\hline PNIPAM-1($40^o$C) & 0.69$\pm $0.03 & 0.87$\pm $0.04 & 0.856$\pm$0.009 \\
\hline PNIPAM-0($40^o$C) & 0.66$\pm $0.01 & 0.80$\pm $0.02 & 0.81$\pm$0.01 \\
\hline PNIPAM-0($34^o$C) & 0.54$\pm $0.02 & 1.13$\pm $0.03 & 1.04$\pm$0.03 \\
\hline
\end{tabular}
 \makeatletter\def\@captype{table}\makeatother
\caption{Values of the dimensionless parameters of $\rho$, ${\left\langle {R_g}^2\right\rangle}^{1/2}_{Zimm}
 /\left\langle R_{s}\right\rangle$ and ${\left\langle {R_g}^2\right\rangle}^{1/2}_{cal}/
 \left\langle R_{s}\right\rangle $.}
\label{table8}
\end{center}

\noindent  Based on the discussion above, the size distribution obtained using the SLS technique is 
consistent with that measured using the TEM technique and can be more accurate to represent the
size distribution of nano-particles in dispersion. The data of
 $g^{\left( 2\right) }\left( \tau \right) $ measured using the DLS technique are determined by the optical and hydrodynamic
properties and size distribution of nano-particles in dispersion together. Therefore the apparent 
hydrodynamic radius (optical weight hydrodynamic radius) can have a large difference with the corresponding mean static radius and the ratio of $R_{app,h}/\left\langle R_{s}\right\rangle$ can 
be larger than 2.

\section{CONCLUSION}

Eq. \ref{mainfit} provides a new method to measure the size distribution of nano-particles in dispersion using the SLS technique. The size distributions of polystyrene latex samples obtained using the TEM and SLS techniques are consistent 
to reveal that the static size distribution is corresponding to the ordinary number particle size distribution. Using the SLS technique to study the scattered light of different shape nano-particles in dispersion further not only can measure the size distribution of nano-particles accurately but also can provide a method to understand the light scattering characteristics based on the knowledge of the structure of nano-particles in dispersion. 
 
\noindent  The data of $g^{\left( 2\right) }\left( \tau \right) $ measured using the DLS technique contains the information of the optical and hydrodynamic
properties of nano-particles together and can be affected largely by the size distribution, therefore the accurate size distribution cannot be obtained from the experimental data of $g^{\left( 2\right) }\left( \tau \right) $ for an
unknown sample. However 
based on the particle size information obtained using the SLS technique, a
possible way to explore the hydrodynamic characteristics of nano-particles in dispersion further is provided.

\end{document}